\RequirePackage{fix-cm}
\documentclass[smallcondensed]{svjour3}     
\smartqed  
\usepackage{graphicx}
\usepackage{amssymb,multirow,mathrsfs,subcaption}
\usepackage{amsmath}
\usepackage{array}
\usepackage{algorithm}
\usepackage{algorithmicx}
\usepackage[noend]{algpseudocode}
\usepackage{paralist}
\usepackage{hyperref}
\usepackage{epsfig}
\usepackage{color,soul,cancel}
\numberwithin{equation}{section}
\usepackage{bigints}
\allowdisplaybreaks

\begin{document}

\title{Two-grid method on unstructured tetrahedra: Applying computational geometry to staggered solution of coupled flow and mechanics problems}



\author{Saumik Dana         \and
        Xiaoxi Zhao  \and
        Birendra Jha
}


\institute{Saumik Dana \at
           University of Southern California, Los Angeles, CA 90007 \\
           \email{sdana@usc.edu}             \\
          \and
          Xiaoxi Zhao \at
          University of Southern California, Los Angeles, CA 90007 \\
          \email{xiaoxiz@usc.edu}             \\
          \and
           Birendra Jha \at
           University of Southern California, Los Angeles, CA 90007 \\
           \email{bjha@usc.edu}             \\
}

\date{Received: date / Accepted: date}

\maketitle

\begin{abstract}
We develop a computational framework that leverages the features of sophisticated software tools and numerics to tackle some of the pressing issues in the realm of earth sciences. The algorithms to handle the physics of multiphase flow, concomitant geomechanics all the way to the earth's surface and the complex geometries of field cases with surfaces of discontinuity are stacked on top of each other in a modular fashion which allows for easy use to the end user. The current focus of the framework is to provide the user with tools for assessing seismic risks associated with energy technologies as well as for use in generating forward simulations in inversion analysis from data obtained using GPS and InSAR. In this work, we focus on one critical aspect in the development of the framework: the use of computational geometry in a two-grid method for unstructured tetrahedral meshes
\end{abstract}
\section{Introduction and motivation}\label{intro}
\begin{figure}[h]
\centering
    \includegraphics[scale=0.4]{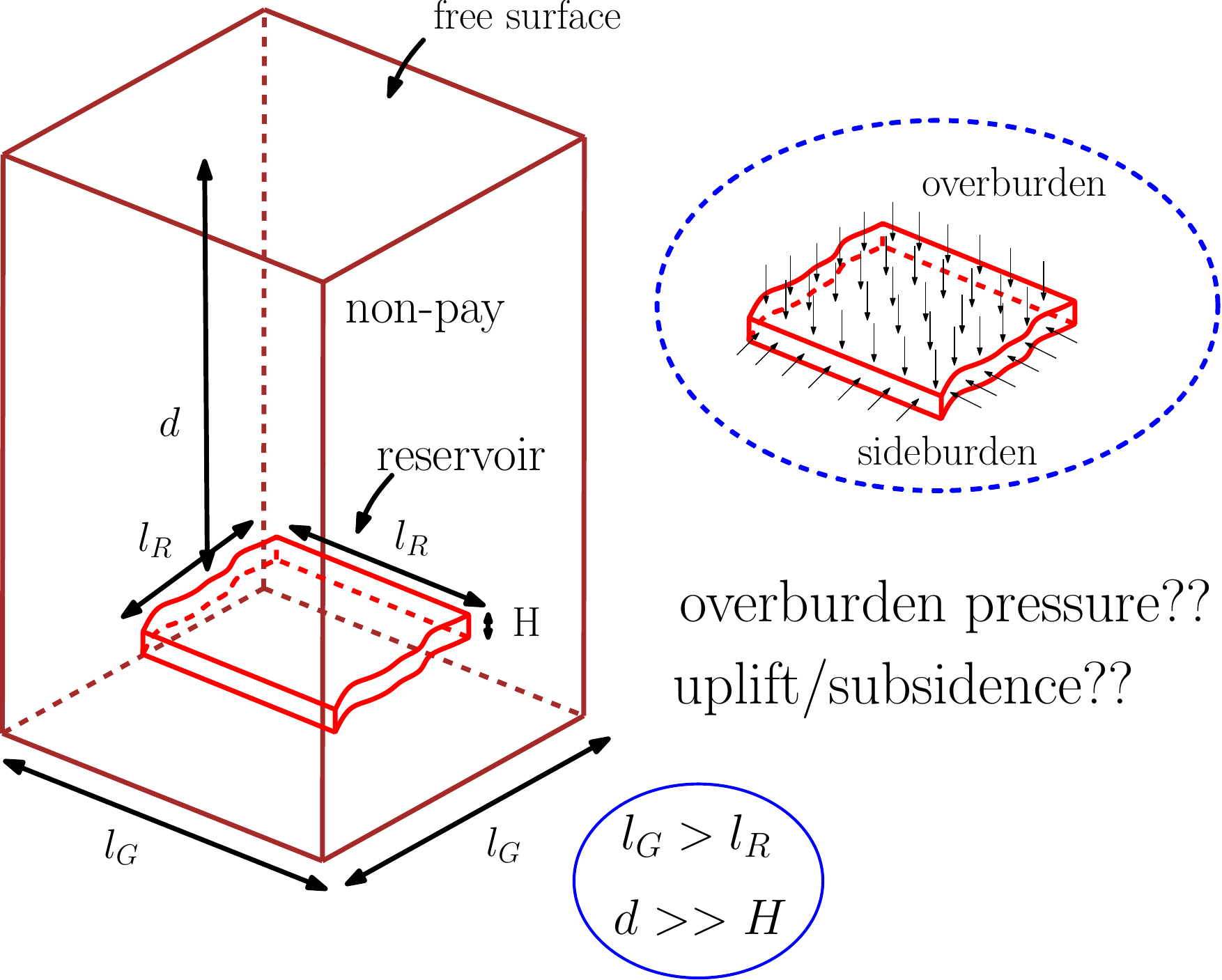}
    \caption{Direct imposition of heuristic overburden pressures on the flow domain completely disregards the mechanical behavior of the surrounding
    rock and obviates a study of fault slip away from the reservoir as well as deformation of earth's surface}
    \label{sketch1}
  \end{figure}

\begin{figure}[h]
    \centering
\includegraphics[scale=0.4]{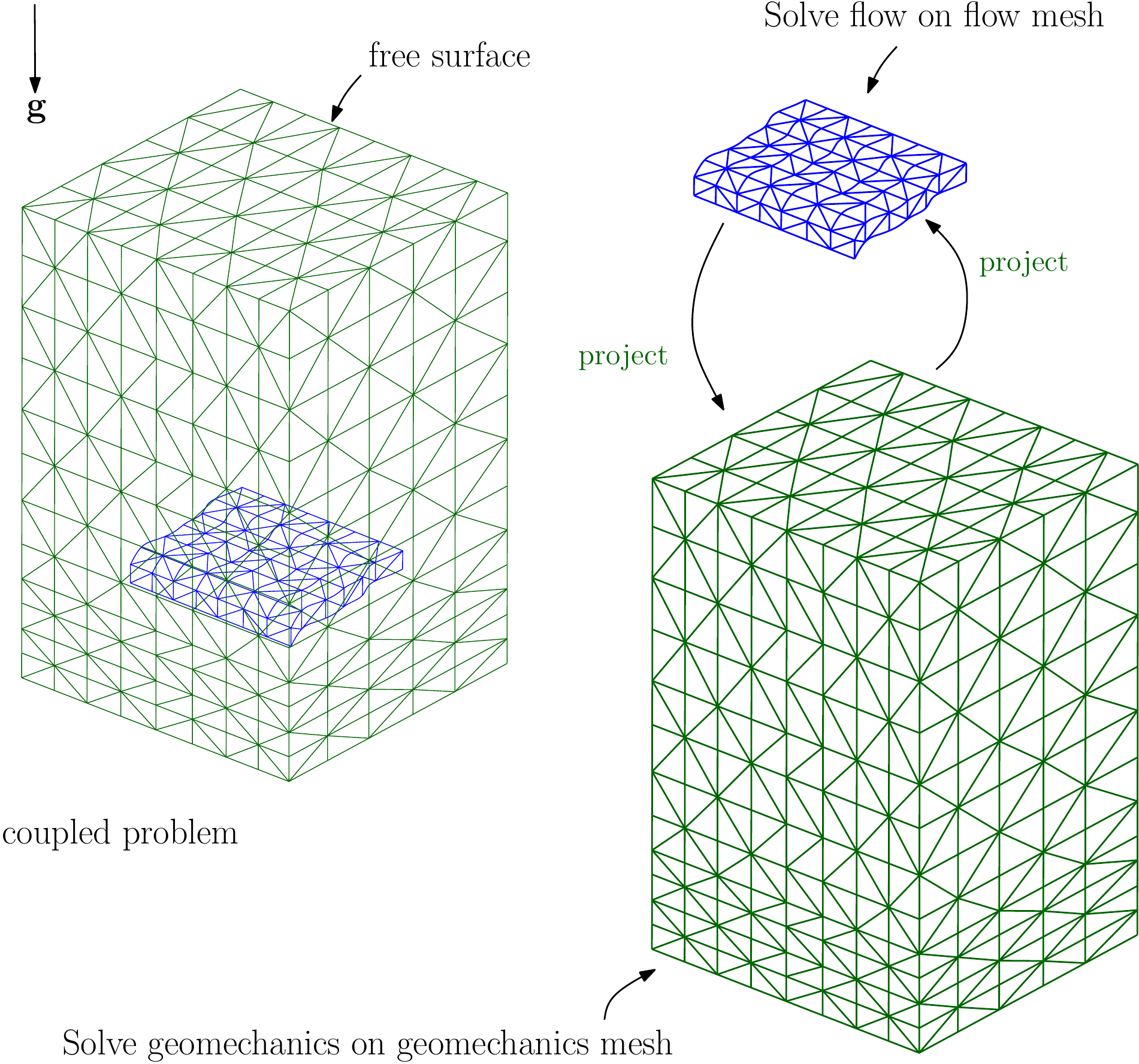}
\caption{The two-grid method enables the study of induced seismicity in
  faults away from reservoir, takes into account the mechanical behavior of
  surrounding rock and allows for determination of surface deformation due to subsurface pressure perturbation}
\label{sketch2}
\end{figure}

\begin{figure}[h]
    \centering
\includegraphics[scale=0.45]{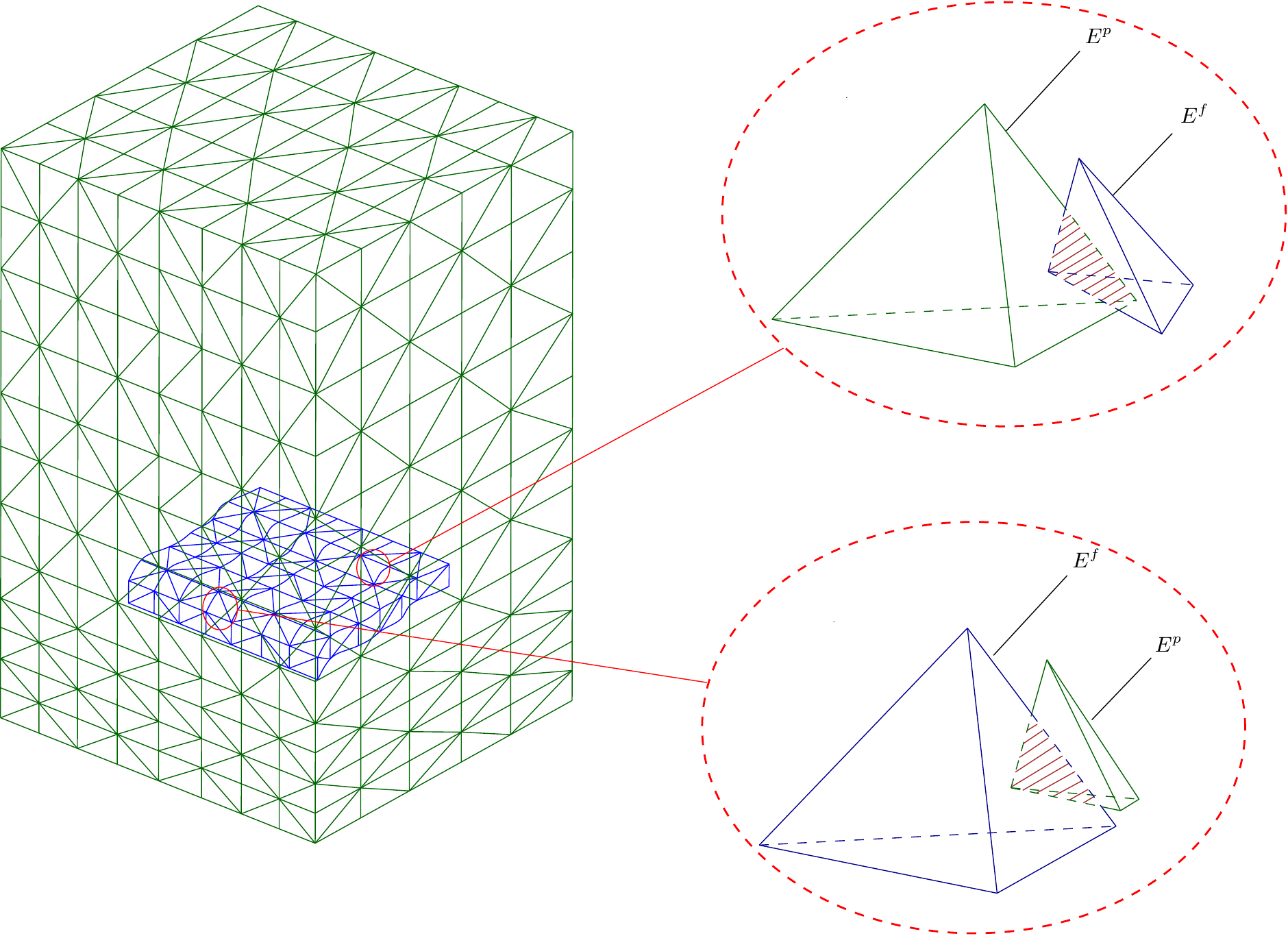}
 \caption{Depiction of couple of intersecting flow-geomechanics element pairs. We refer to the flow element as $E^f$ and geomechanics element as $E^p$. Depiction is to reiterate the point that there is no restriction of whether the flow or geomechanics element needs to be smaller than the other}
\label{explain}
\end{figure}

\noindent The development of a computationally inexpensive framework with the capability to model deformation of the earth's surface and fault slip due to deep subsurface pressure perturbations associated with multiphase flow is critical from two different standpoints
\begin{itemize}
\item Assessing the seismic risks associated with carbon sequestration, enhanced geothermal systems, waste water disposal, enhanced oil recovery thereby offering guiding protocols for the design of these operations \cite{induced1,induced2,induced3,one,two,three,geo1,geo2} 
\item Estimating subsurface properties using inversion analysis of ground deformation data obtained from Global Positioning System (GPS) and interferometric synthetic
aperture radar (InSAR). This typically requires multiple forward simulations with the geomechanical domain extending all the way to the earth's surface \cite{1,2,3,4,5,6,7,8,9,jha2015reservoir}
\end{itemize}
The major bugbear to developing such a framework is the computationally intractable size of the geomechanical grid no matter which numerical method is used to resolve the coupled system of equations. The typical approaches to avoid such issues are to impose an overburden pressure directly on the
reservoir thus treating it as a coupled problem domain (see Fig. \ref{sketch1}) or to model flow on huge domain with zero permeability cells mimicking the no flow boundary condition between the flow and non-flow region. The former approach precludes a study of surface deformation, does not mimic the true effect of the overburden on the stress sensitive reservoir, and is incapable of capturing induced seismicity inside faults away from reservoir whereas the latter approach is computationally intractable for large field scale problems due to memory requirements. In order to address this, we develop a two-grid
coupled multiphase flow and geomechanics framework which allows for spatial decoupling of the flow and geomechanics domains with the geomechanics subproblem being resolved on a separate grid with a larger spatial extent going all the way to the free surface (see Fig. \ref{sketch2}). This computational framework is built on top of a staggered solution algorithm that solves the flow and mechanics subproblems sequentially and iteratively.
\par
Typically, in such problems, the geomechanics mesh is expected to be coarser than the flow mesh everywhere the two meshes exist, but we generalize the method to the cases where the geomechanics elements can be smaller than the flow elements in small localized region where capturing the mechanics is more pertinent. Furthermore, we would like to generalize the notion of the two-grid from structured hexahedral meshes (as was done in \cite{dana-2018}) to unstructured tetrahedral meshes. A depiction of the intersection of two tetrahedral elements in given in Fig. \ref{explain}. In this document, we focus on demonstrating the convergence of the two-grid method for unstructured tetrahedral grids using the classical Mandel's problem \cite{mandel-1953,ref27} analytical solution. We work two cases, one in which we have a finer mesh for flow, and the other in which we have a finer mesh for mechanics.
\par 
This paper is structured as follows: the governing equations are provided in section 2, the solution strategy is explained in section 3, the salient features of our software are elucidated in section 4, the numerical simulations for Mandel's problem are provided in section 5, and the conclusions and outlook are provided in section 6.
\section{Governing equations}\label{sec:goveq}
We use a classical continuum representation in which the fluids and the solid skeleton are viewed as overlapping continua \cite{BeaJ1972,CouO2005}.
The governing equations for coupled flow and geomechanics are obtained from conservation of mass and balance of linear momentum. We assume that the deformations are small, that the geomaterial is isotropic, and that the conditions are isothermal. 

Under the quasistatic assumption for earth displacements, the governing equation for linear momentum balance of the solid/fluid system can be expressed as
\begin{equation}\label{e:linmom}
\nabla\cdot\boldsymbol{\sigma}+\rho_b\boldsymbol{g}=\boldsymbol{0},
\end{equation}
where~$\boldsymbol{\sigma}$ is the Cauchy total stress tensor, $\boldsymbol{g}$ is the gravity vector, and $\rho_b=\phi\sum_{\beta}^{n_{\text{phase}}}\rho_{\beta}S_{\beta}+(1-\phi)\rho_s,$ is the bulk density, $\rho_{\beta}$ and $S_{\beta}$ are the density and saturation of fluid phase~$\beta$, $\rho_s$ is the density of the solid phase, $\phi$ is the true porosity, and~$n_{\text{phase}}$ is the number of fluid phases. The true porosity is defined as the ratio of the pore volume $V_p$ to the bulk volume $V_b$ in the current (deformed) configuration.

Assuming that the fluids are immiscible, the mass-conservation equation for each phase $\alpha$ is
\begin{equation}\label{e:masscons_mp}
\frac{d m_{\alpha}}{d t}+\nabla\cdot{\boldsymbol{w}}_{\alpha}=\rho_{\alpha} f_{\alpha},
\end{equation}
where the accumulation term ${d m_{\alpha}}/{d t}$ describes the time variation of fluid mass relative to the motion of the solid skeleton, ${\boldsymbol{w}}_{\alpha}$~is the mass-flux of fluid phase $\alpha$ relative to the solid skeleton, and $f_{\alpha}$ is the volumetric source term for phase~$\alpha$. Balance equations~\eqref{e:linmom} and~\eqref{e:masscons_mp} are coupled by virtue of poromechanics. On one hand, changes in the pore fluid pressure lead to changes in effective stress, and induce deformation of the porous material---such as ground subsidence caused by groundwater withdrawal. On the other hand, deformation of the porous medium affects fluid mass content and fluid pressure. 
The simplest model of this two-way coupling is Biot's macroscopic theory of poroelasticity \cite{BioM1941,GeeJ1957,CouO1995}. The specific equations of single phase poromechanics are provided in Appendix \ref{spporo}
\section{Solution strategy}
\begin{figure}[h]
\centering
\includegraphics[scale=0.9]{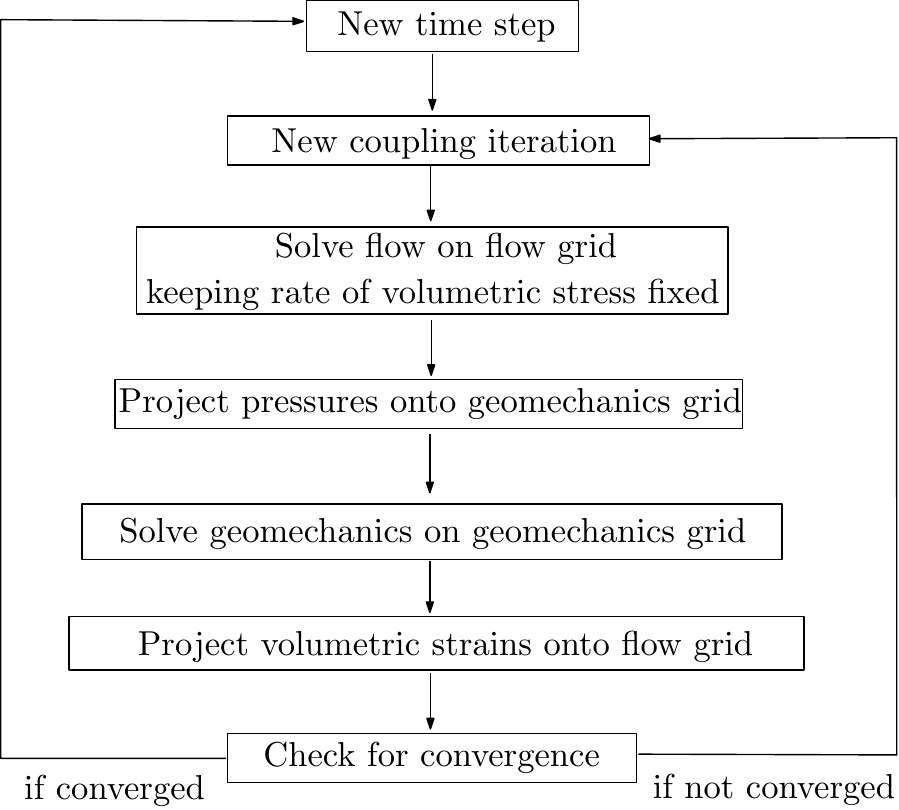}
\caption{Two grid staggered solution algorithm}
\label{sketch3}
\end{figure}
\noindent The two-grid approach is built on top of a staggered
solution algorithm (Fig. \ref{sketch3}) in which the flow and geomechanics subproblems are solved
sequentially and iteratively using a fixed stress split iterative scheme
\cite{jha2014coupled,kim-stability,ref12,mikelic-2014,nicola-fixedstress,white-block,scalable,danacg,danacmame,dana2020,almani,almanicg,both2019,white2019,borregales2019,storvik2019,radu-robust,raduadaptive,lu_wheeler}.
\subsection{Space and time discretization}
We use the finite volume method for the discretization of the flow problem \cite{AziK1979}, and the nodal-based finite element method for the discretization of the mechanics problem \cite{HugT1987,ZieO2005}. 
This space discretization is locally mass conservative at the element level, and enjoys excellent stability properties \cite{JhaB2007,PhiP2007-01,PhiP2007-02,KimJ2011-spej}.
The pressures and saturations degrees of freedom are located at the element centers of the flow grid, and the displacement vector degrees of freedom are located at the element nodes of the mechanics grid. 
In quasi-static poromechanics, the time derivative appears only in the accumulation term of the fluid mass balance equation, and we treat it using a fully implicit Backward Euler time integration scheme.
\subsection{Two-grid operators}
\begin{algorithm}[h]
\begin{algorithmic} 
\caption{Determining the relationship between flow-geomechanics element pairs}
\label{ifw} 
\For{Loop over all geomechanics elements}
\For{Loop over all flow elements}
\State count1 $\gets$ 0
\For{Loop over four vertices of flow element}
\If{Vertex inside geomechanics element}
\State count1 $\gets$ count1 + 1
\EndIf
\EndFor
\If{count1!=0}
\State Flow to geomechanics projection operator $\gets Meas(E^f)$ 
\State Geomechanics to flow projection operator $\gets Meas(E^p)$ 
\EndIf
\State count2 $\gets$ 0
\For{Loop over four vertices of geomechanics element}
\If{Vertex inside flow element}
\State count2 $\gets$ count2 + 1
\EndIf
\EndFor
\If{count2!=0}
\State Flow to geomechanics projection operator $\gets Meas(E^f)$ 
\State Geomechanics to flow projection operator $\gets Meas(E^p)$ 
\EndIf
\EndFor
\EndFor
\end{algorithmic}
\end{algorithm} 

\noindent The basic idea of the two-grid approach is to transfer pore pressures from the flow domain to the geomechanics domain and conversely transfer volumetric strain from the geomechanical domain to the flow domain. The transfered pore pressure at every geomechanics element is a volume average of
the pore pressures at the flow elements interacting with the geomechanics element. Likewise, the transfered volumetric strain at each flow element is a volume
average of the volumetric strains at the geomechanics elements interacting with the flow element. The global two-grid
operators are constructed in the pre-processing step by applying a local two-grid procedure to each pair of intersecting finite elements of the two
grids as shown in Algorithm \ref{ifw}. The relationship between the flow-geomechanics element pair is obtained by determining the barycentric coordinates of the vertices of the flow element with respect to the geomechanics element and vice versa as explained in Algorithm \ref{ifw}. 
\section{Software implementation}
\begin{figure}[h]
\centering
\includegraphics[scale=0.7]{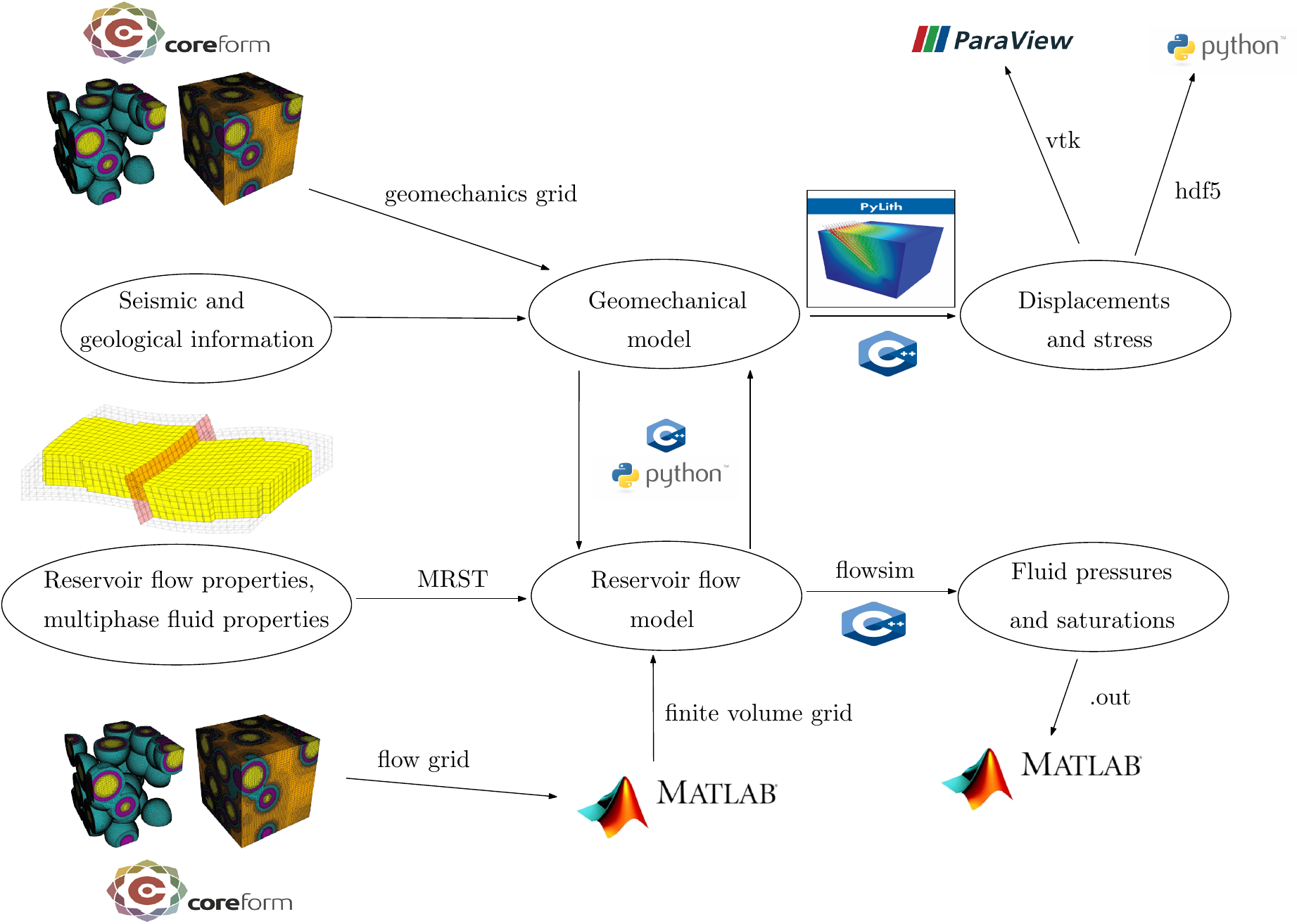}
\caption{Workflow}
\label{nutshell}
 \end{figure}
We developed a coupled multiphase flow and geomechanical simulator by integrating our version of Stanford's General Purpose Research Simulator (GPRS) \cite{CaoH2002,GPRSMan2010} called flowsim which operates as the flow simulator into an open source framework called PyLith \cite{pylithman2011,AaaB2013} which operates as the mechanics simulator. The workflow of our software are elucidated in Fig. \ref{nutshell}. Available seismic and
geological information are used to construct a geomechanical model and populate it with poromechanical properties.
Reservoir flow properties (porosity and permeability) and fluid properties (fluid density, viscosity, and compressibility)
are obtained from an older uncoupled model and transferred to the geomechanical model using the Matlab Reservoir
Simulation Toolbox \cite{lie2019}.
\subsection{Flow Simulator}
flowsim is a general purpose, object-oriented, field-scale simulator for multiphase (oil, water, and gas) and multicomponent (e.g. methane, ethane, decane, CO$_2$, N$_2$, and water) flow through subsurface. It treats element connections through a general connection list, which allows for both structured and
unstructured grids. It is capable of handling complex production and injection scenarios in the field,
such as wells perforated at multiple depths and flowing under variable rate and pressure controls.
\subsection{Mechanics Simulator} 
\begin{figure}[h]
\centering
\includegraphics[scale=0.25]{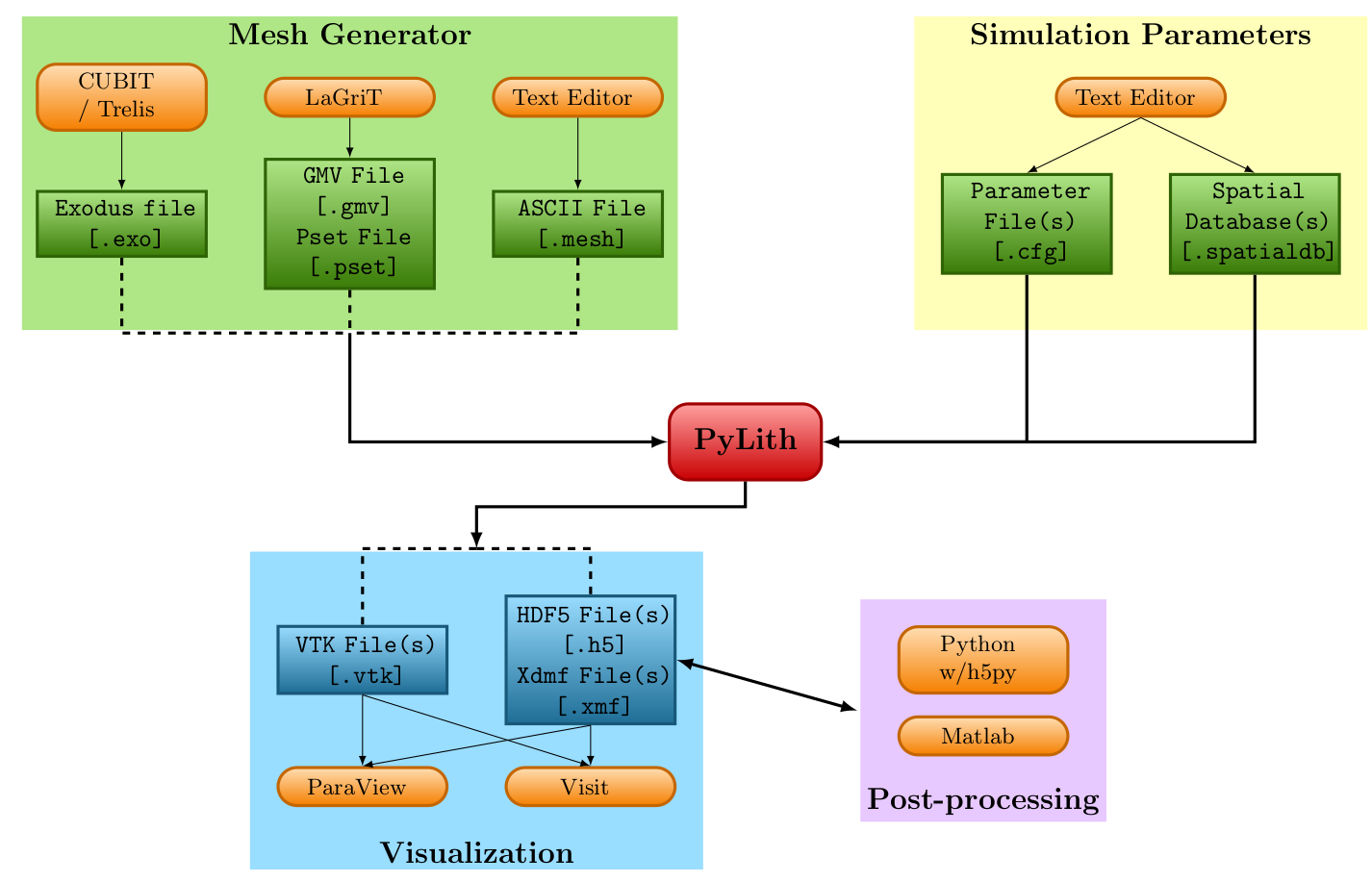}
\caption{PyLith requires a finite-element mesh, simulation parameters, and spatial databases (defining the spatial variation of various parameters). It writes the solution
output to either VTK or HDF5/Xdmf files, which can be visualized with ParaView or Visit. Post-processing is generally done
using the HDF5 files with Python or Matlab scripts \cite{aagaard1}}
\label{pynut}
\end{figure}
\noindent PyLith is a open-source finite element code for the simulation of static and dynamic
large-scale deformation problems \cite{aagaard1,aagaard2}. It uses an implicit formulation to solve quasi-static problems and an explicit formulation to solve dynamic rupture problems. The salient features of Pylith are elucidated in Fig. \ref{pynut}. Some of the advantages of PyLith are: (a) It is written using C++ and Python languages and is extendable, (b) it is suitable for parallel computing, (c) it allows localized deformation along discrete features, such as faults, (d) it is well integrated with meshing codes, such as LaGriT for tetrahedral meshes \cite{lagrit} and CUBIT for both tetrahedral and hexahedral meshes \cite{cubit}
\subsection{Gridding}
We employ different grids for flowsim and PyLith to go with our two-grid method. The finite element grids are generated using CUBIT \cite{cubit} mesh generation software and exported in the Exodus-II format. Since the numerical method employed for the flow model is finite volume instead of finite element, we employ some further processing of the flow grid file to render it for the finite volume calculations. We process the Exodus-II grid file using a MATLAB script to generate the equivalent
finite volume grid in the domain with element centroid coordinates, element bulk volumes, and face transmissibilities in the Corner Point Geometry format \cite{eclipse}
\subsection{Coupling}
We design a C++ class, iflowsim, to allow communication between the
flow and the mechanics simulators. The two-grid operators built on top of iflowsim to project information between the two grids. PyLith supports distributed memory parallelization (Message Passing Interface or MPI) whereas flowsim’s parallelization is based on the shared memory architecture (Multiprocessing or OpenMP)

\section{Mandel's problem}
\begin{figure}[h]
\centering
\includegraphics[scale=0.8]{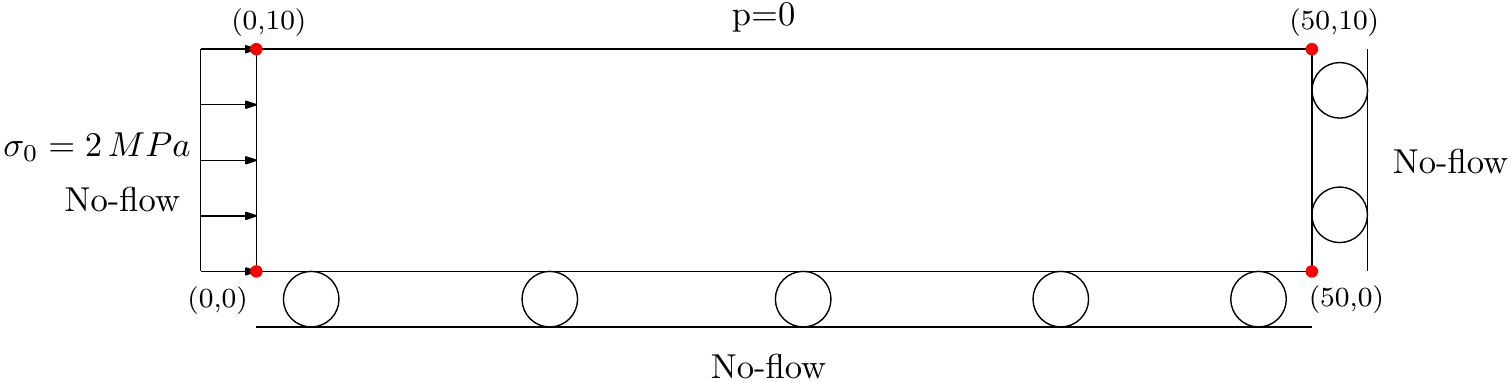}
\caption{Schematic for Mandel's problem}
\label{mandelfig1}
\end{figure}

\begin{figure}[h]
\begin{subfigure}{.5\textwidth}
\centering
\includegraphics[scale=0.25]{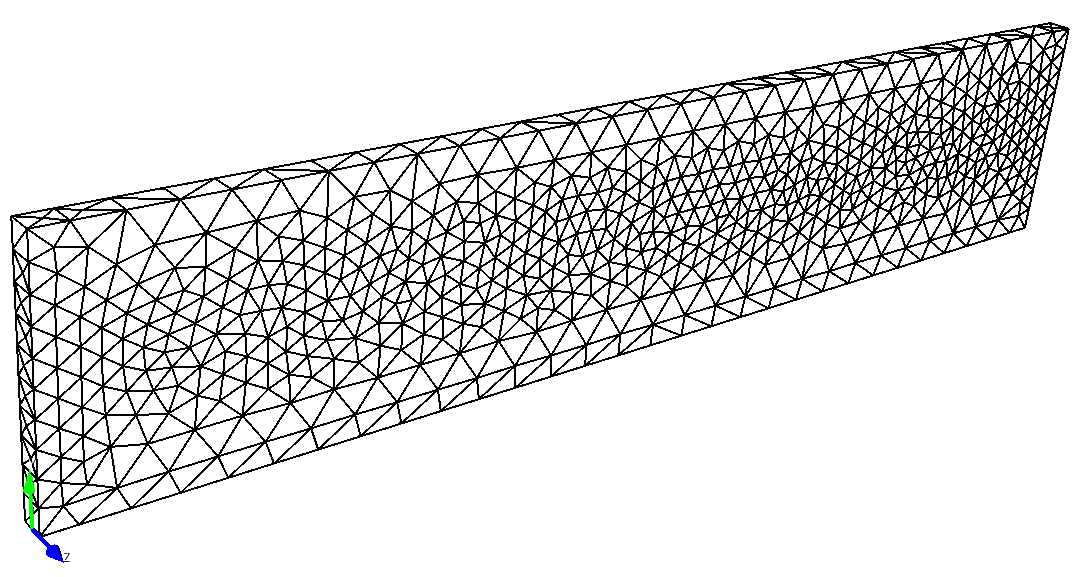}
\caption{Flow mesh, 5706 elements}
\end{subfigure}
\begin{subfigure}{.5\textwidth}
\centering
\includegraphics[scale=0.25]{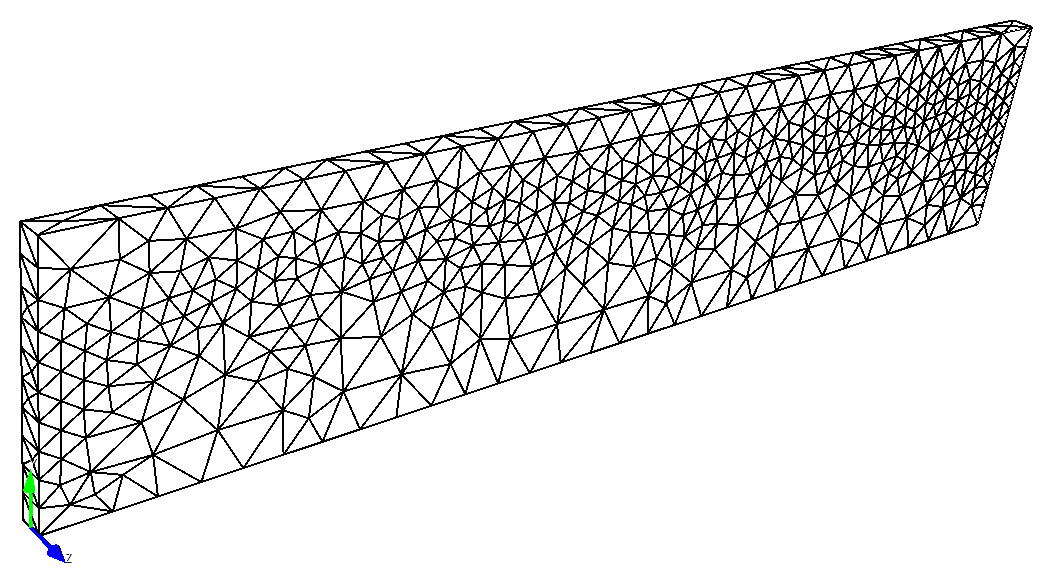}
\caption{Pylith mesh, 3697 elements}
\end{subfigure}
\caption{Finer mesh for flow compared to Pylith}
\label{mandelfig2}
\end{figure}

\begin{figure}[h]
\begin{subfigure}{.5\textwidth}
\centering
\includegraphics[scale=0.25]{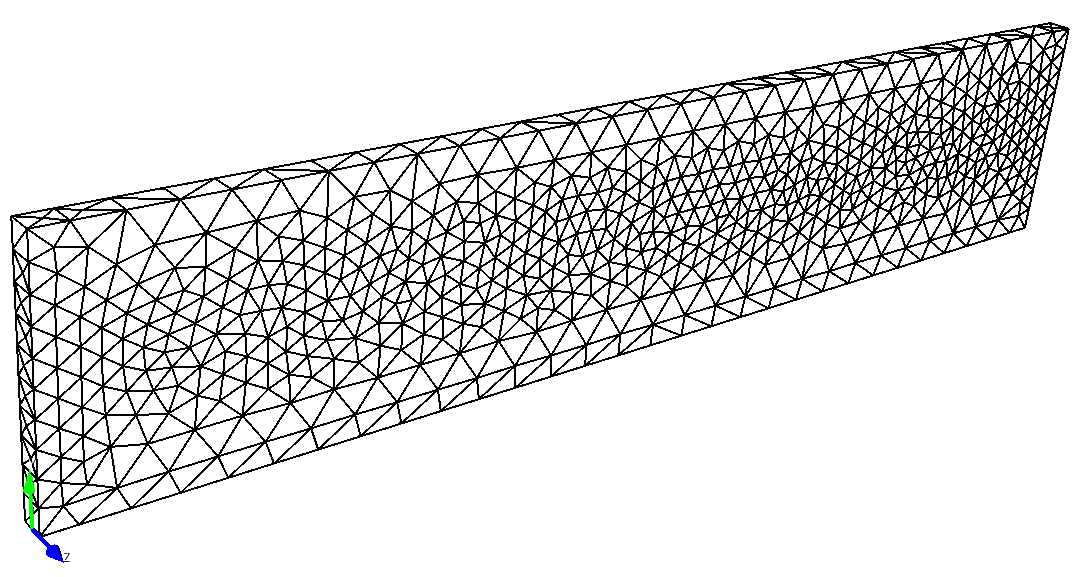}
\caption{Pylith mesh, 5706 elements}
\end{subfigure}
\begin{subfigure}{.5\textwidth}
\centering
\includegraphics[scale=0.25]{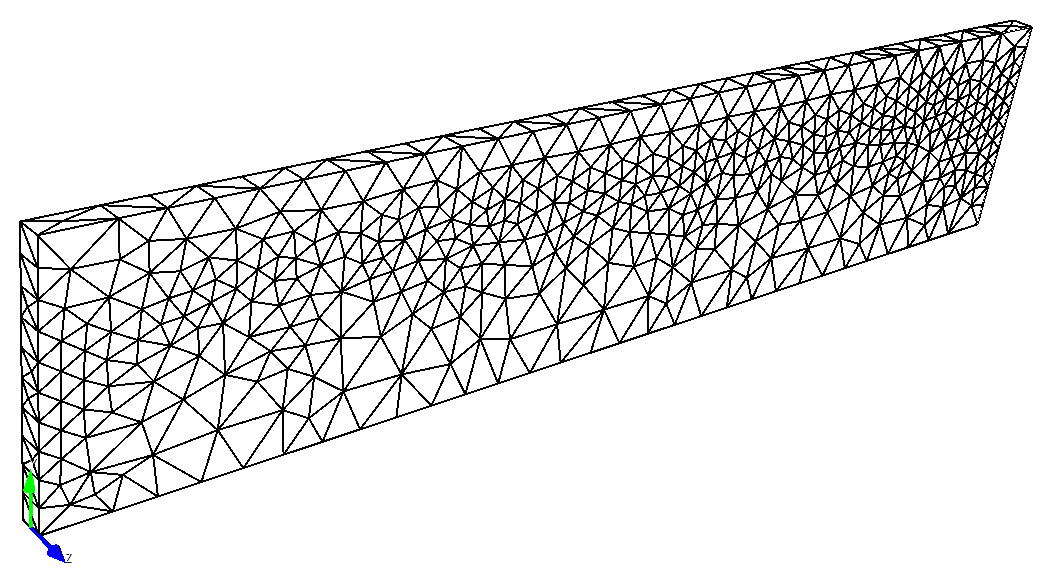}
\caption{Flow mesh, 3697 elements}
\end{subfigure}
\caption{Finer mesh for Pylith compared to flow}
\label{mandelfig3}
\end{figure}

\begin{figure}[H]
\begin{subfigure}{\textwidth}
\centering
\includegraphics[trim={2cm 8.5cm 2cm 8.5cm},clip,scale=0.75]{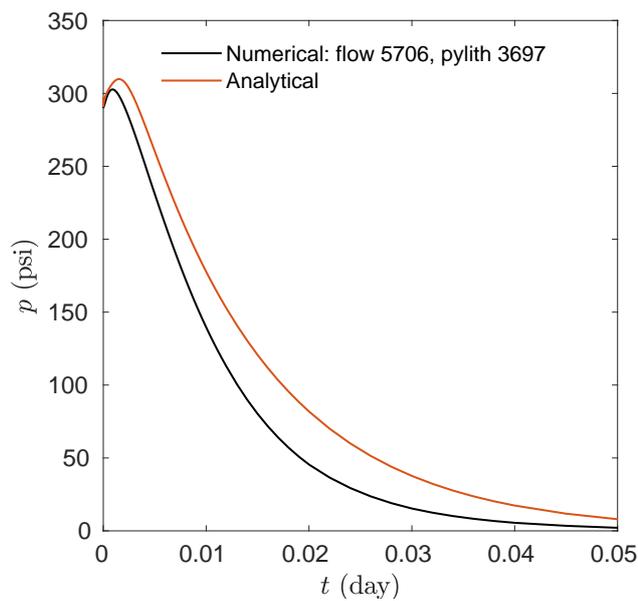}
\caption{Flow mesh: 5706 elements and pylith mesh: 3697 elements}
\end{subfigure}
\begin{subfigure}{\textwidth}
\centering
\includegraphics[trim={2cm 8.5cm 2cm 8.5cm},clip,scale=0.75]{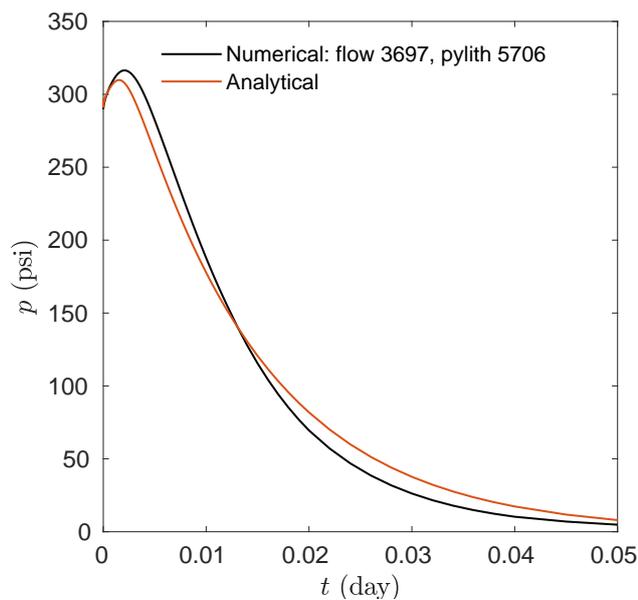}
\caption{Flow mesh: 3697 elements and pylith mesh: 5706 elements }
\end{subfigure}
\caption{Comparison of non-monotonic pore pressure response}
\label{mandelfig4}
\end{figure}

\noindent The Mandel’s problem \cite{mandel-1953,ref27} has been used as a benchmark problem for testing the validity of numerical codes of coupled poroelasticity. Its main feature, the Mandel-Cryer effect, is that the pore pressure at the center of a loaded specimen rises above its initial value because of the two-way coupling between fluid flow and solid deformation. It involves a long specimen of rectangular cross section pressed on one side with an impermeable plate that applies a constant compressive stress and fixed on two sides using impermeable roller boundaries (Fig. \ref{mandelfig1}). The fourth side of the cross section is free from normal and shear stresses (traction-free boundary) and is open to the atmosphere (constant pressure boundary). The porous medium is saturated with a slightly compressible fluid, water, with initial pressure set at the reference value. Since the specimen is long, we assume plane strain conditions, namely, that the displacement and fluid flux vanish in the z direction (perpendicular to the 2-D domain). 

With these boundary conditions, the
three-dimensional equations of
poroelasticity reduce to one-dimensional equations for $\sigma_{xx}(y,t)$ and $p(y,t)$, which can be solved analytically \cite{mandel-1953,ref27}. At $t=0^+$ , a uniform undrained pressure is generated by the Skempton effect, along with uniform stress $\sigma_{xx}=-\sigma_0$. The specimen expands toward the top boundary due to the Poisson effect. As time
progresses, the pressure near the top boundary decreases because of fluid drainage, which makes the specimen more compliant there. If the hydraulic diffusivity is small, the effect of drainage is not observed immediately near the no-flux bottom boundary. This results into load transfer of compressive total stress toward the bottom boundary, in response to which the pressure there continues to rise above its undrained value. At long times, all excess pressure vanishes and a uniform horizontal stress, $\sigma_{xx}=-\sigma_0$, returns. Hence, the
pressure evolution at points away from the drained boundary is nonmonotonic, a phenomenon not
observed in a purely diffusive process such as that modeled by the Terzaghi theory, where the pressure is uncoupled from the solid deformation.

In one simulation, we employ a finer mesh for the flow problem as shown in Fig. \ref{mandelfig2} whereas in the other simulation, we employ a finer mesh for the mechanics problem as shown in Fig. \ref{mandelfig3}. Fig. \ref{mandelfig4} compares the pressure response at the right bottom corner $(x=50,y=0)$ at different times.

\section{Conclusions and outlook}
We developed a two-grid method which leverages computational geometry on unstructured tetrahedral meshes that allows us to solve the flow and geomechanics subproblems on separate grids. This allows use for flexibility in the choice of the two separate meshes depending on the needs of the specific problem. There is no restriction on whether the flow element needs to be finer or coarser than the geomechanics element in any region of the subdomains. This method also allows us to have separate domains for flow and geomechanics, which is critical for large scale field problems typically encountered in energy technologies. We shall now employ our computational framework to solve field scale problems with faults in future articles.

\begin{acknowledgements}
The first author would like to thank Sreekanth Arikatla (Staff R\&D Engineer, Kitware Inc.) for discussions on computational geometry for unstructured tetrahedra
\end{acknowledgements}

%
\section*{Conflict of interest}

The authors declare that they have no conflict of interest.

\appendix
\section{Single-phase poromechanics}\label{spporo}
For isothermal single-phase flow of a slightly compressible fluid in a poroelastic medium with no stress dependence of permeability, the single-phase fluid mass conservation equation reduces to
\begin{equation}\label{e:masscons_sp}
\frac{d m}{d t}+\nabla\cdot \boldsymbol{w}=\rho_f f,
\end{equation}
where $m$~is the fluid mass content (fluid mass per unit bulk volume of porous medium), $\rho_f$~is the fluid density, $\boldsymbol{w}=\rho_f\boldsymbol{v}$ is the fluid mass flux (fluid mass flow rate per unit area and time), and $\boldsymbol{v}$~is the seepage velocity relative to the deforming skeleton, given by Darcy's law:
\begin{equation}\label{e:darcy_sp}
\boldsymbol{v}=-\frac{\boldsymbol{k}}{\mu}\left(\nabla p-\rho_f\boldsymbol{g} \right),
\end{equation}
where $\boldsymbol{k}$~is the intrinsic permeability tensor, $\mu$~is the fluid dynamic viscosity and $p$~is the pore-fluid pressure \cite{BeaJ1972}.
It is useful to define the fluid content variation $\zeta$,
\begin{equation}
	\zeta:=\frac{\delta m}{\rho_{f,0}},
\end{equation}
where $\delta m=m-m_0$ is the increment in fluid mass content with respect to the initial reference state, and $\rho_{f,0}$~is the reference fluid density. The self-consistent theory of poroelastic behavior proposed by \cite{BioM1941} links the changes in total stress and fluid pressure with changes in strain and fluid content. Following \cite{CouO1995}, the poroelasticity equations can be written in incremental form as
\begin{equation}\label{e:stressstrain}
\begin{split}
\delta \boldsymbol{\sigma}&=\boldsymbol{C}_{dr}:\boldsymbol{\varepsilon}-b\delta p\boldsymbol{1}\\
\zeta&=b\varepsilon_v + \frac{1}{M}\delta p,
\end{split}
\end{equation}
where $\boldsymbol{C}_{dr}$ is the rank-4 drained elasticity tensor, $\boldsymbol{1}$~is the rank-2 identity tensor, $\boldsymbol{\varepsilon}$~is the linearized strain tensor, defined as the symmetric gradient of the displacement vector~$\boldsymbol{u}$,
\begin{equation}
	\boldsymbol{\varepsilon}:= \frac{1}{2}\left(\nabla\boldsymbol{u}+\nabla^T\boldsymbol{u}\right),
\end{equation}
and $\varepsilon_v=\text{tr}(\boldsymbol{\varepsilon})$ is the volumetric strain. Note that we use the convention that tensile stress is positive. It is useful to express the strain tensor as the sum of its volumetric and deviatoric components:
\begin{equation}
	\boldsymbol{\varepsilon}=\frac{1}{3}\varepsilon_v\boldsymbol{1} +\boldsymbol{e},
\end{equation}
from which it follows that the volumetric stress $\sigma_v=\text{tr}(\boldsymbol{\sigma})/3$ satisfies:
\begin{equation}
	\delta\sigma_v=K_{dr}\varepsilon_v-b\delta p.
\end{equation}
Equation~\eqref{e:stressstrain} implies that the effective stress in single-phase poroelasticity, responsible for skeleton deformation, is defined in incremental form as
\begin{equation}
	\delta\boldsymbol{\sigma}':=\delta\boldsymbol{\sigma}
	+b\delta p\boldsymbol{1}.
\end{equation}
Biot's theory of poroelasticity has two coupling coefficients: the Biot modulus~$M$ and the Biot coefficient~$b$. They are related to rock and fluid properties as \cite{CouO1995}
\begin{equation}\label{e:biotmod}
\frac{1}{M}=\phi_0 c_f+\frac{b-\phi_0}{K_s}, \quad b=1-\frac{K_{dr}}{K_s},
\end{equation}
where $c_f=1/K_f$ is the fluid compressibility, $K_f$ is the bulk modulus of the fluid, $K_s$ is the bulk modulus of the solid grain, and $K_{dr}$ is the drained bulk modulus of the porous medium. To set the stage for the numerical solution strategy of the coupled problem, it is useful to write the fluid mass balance equation~\eqref{e:masscons_sp} (the pressure equation) in a way that explicitly recognizes the coupling with mechanical deformation. Equations~\eqref{e:stressstrain} state that the increment in fluid mass content has two components: increment due to expansion of the pore space and increment due to increase in the fluid pressure. Assuming small elastic deformations and applying linearization from the reference state to the current state, we can write Eqs.~\eqref{e:stressstrain} as
\begin{align}\label{e:stressstrain1}
\boldsymbol{\sigma}-\boldsymbol{\sigma}_0&=\boldsymbol{C}_{dr}:\boldsymbol{\varepsilon}-b\left(p-p_0\right)\boldsymbol{1},\\\label{e:fluidmass1}
\frac{1}{\rho_{f,0}}(m-m_0)&=b\varepsilon_v + \frac{1}{M}(p - p_0).
\end{align}
Substituting Eq.~\eqref{e:fluidmass1} into Eq.~\eqref{e:masscons_sp}, we obtain the fluid mass balance equation in terms of the pressure and the volumetric strain:
\begin{equation}\label{e:massbal_fixedstrain}
\frac{1}{M}\frac{\partial p}{\partial t}+b\frac{\partial \varepsilon_v}{\partial t}+\nabla\cdot \boldsymbol{v}=f.
\end{equation}
Linearizing the relation between volumetric total stress and volumetric strain with respect to the reference state,
\begin{equation}\label{e:stressstrain2}
\sigma_v-\sigma_{v,0}=K_{dr}\varepsilon_v-b\left(p-p_0\right),
\end{equation}
allows us to express the change in porosity as the sum of a volumetric stress component and a fluid pressure component. From $m=\rho_f\phi$ and Eq.~\eqref{e:fluidmass1},
\begin{equation}\label{e:poro_twoway}
\frac{\rho_f}{\rho_{f,0}}\phi-\phi_0 = \frac{b}{K_{dr}}\left(\sigma_v - \sigma_{v,0}\right)+\left(\frac{b^2}{K_{dr}}+\frac{1}{M}\right)\left(p-p_0\right).
\end{equation}
Using the effective stress equation, Eq.~\eqref{e:stressstrain2}, we can rewrite Eq.~\eqref{e:massbal_fixedstrain} in terms of pressure and volumetric total stress:
\begin{equation}\label{e:massbal_fixedstress}
\left(\frac{b^2}{K_{dr}}+\frac{1}{M}\right)\frac{\partial p}{\partial t}+\frac{b}{K_{dr}}\frac{\partial \sigma_v}{\partial t}+\nabla\cdot \boldsymbol{v}=f.
\end{equation}
\bibliographystyle{spmpsci}      
\bibliography{references}   



\end{document}